\documentclass[twocolumn,preprintnumbers,pre]{revtex4-1}
\usepackage{graphicx, subfigure}
\usepackage{amssymb, amsmath,amssymb,amsfonts}
\usepackage{amsthm,mathrsfs,amsopn}
\usepackage{dcolumn}
\usepackage{bm}
\usepackage{enumerate}
\usepackage[normalem]{ulem}
\usepackage{color}
\usepackage[utf8]{inputenc}
\usepackage{mathtools}
\usepackage{tikz}
\usetikzlibrary{arrows.meta,calc}
\usepackage{orcidlink}
\graphicspath{{./Figures}}
\theoremstyle{plain} %% This is the default

\usepackage{float}
\usepackage{wrapfig}
\usepackage{hyperref}
\usepackage{xcolor}
\usepackage{graphicx}
\usepackage{microtype}
\usepackage{comment}
\newcommand{\rowlabelsize}{\fontsize{5}{6}\selectfont}
 
\begin{document}

\title{Traveling chimeras and collective coordination in $\beta$-cell networks}

\author{Carine Simo$^{1,2}$, Venceslas Nguefoue Meli$^{1,2}$, Patrick Louodop$^{1,2}$\,\orcidlink{0000-0002-2975-2420}, Samuel Bowong$^{3,2}$\,\orcidlink{0000-0001-8909-1225}, and Thierry Njougouo$^{4,2}$\,\orcidlink{0000-0001-7706-7674}}
\email{thierry.njougouo@imtlucca.it}
\affiliation{$^1$Research Unit Condensed Matter, Electronics and Signal Processing,\\ University of Dschang, P.O Box 67 Dschang, Cameroon.}
\affiliation{$^2$MoCLiS Research Group, Dschang, Cameroon.}
\affiliation{$^3$Department of Mathematics and Computer Science, Faculty of Science,\\ University of Douala, PO Box 24157 Douala, Cameroon.}
\affiliation{$^4$IMT School for Advanced Studies Lucca\\ Piazza San Francesco 19, 55100 Lucca, Italy.}

\begin{abstract}
Pancreatic $\beta$-cells play a central role in maintaining glucose homeostasis through the pulsatile secretion of insulin. This essential function relies not only on intracellular regulatory mechanisms but also on coordinated interactions among $\beta$-cells within the islets of Langerhans. Disruptions in this intercellular coordination are increasingly implicated in metabolic disorders such as type~II diabetes. In this work, we employ a computational framework to investigate the collective dynamics of a network of coupled $\beta$-cells interacting through a nonlocally coupled ring topology that incorporates both electrical and metabolic coupling pathways. This topology provides an idealized representation of short- and longer-range interactions and allows their dynamical consequences to be studied systematically. Numerical simulations reveal a variety of emergent behaviors, including synchronization, traveling waves, and traveling chimera states, in which coherent and incoherent domains coexist and propagate across the network. These findings provide new insight into the mechanisms governing coordinated $\beta$-cell activity and the regulation 
of pulsatile insulin secretion. By clarifying how coupling structure and intercellular communication shape islet-wide dynamics, this work contributes to a deeper understanding of the dysfunctions underlying diabetes.
\end{abstract}

\maketitle

\section{Introduction}

Maintaining glucose homeostasis depends on a balance among intestinal glucose absorption, hepatic glucose production, and glucose release into the bloodstream~\cite{roder2016pancreatic,rorsman2018pancreatic}. At the heart of this process are the pancreatic $\beta$ cells, located within the islets of Langerhans, which represent approximately 60--80\% of the islet cell population~\cite{Dunne2024_PancreaticBetaCells}. These specialized cells are responsible for the synthesis, storage, and release of insulin---a pivotal hormone in glucose regulation. In response to elevated blood glucose, they secrete insulin to facilitate glucose uptake and storage in peripheral tissues, thereby restoring 
normoglycemia~\cite{rorsman2018pancreatic,Dunne2024_PancreaticBetaCells}.

Upon glucose stimulation, $\beta$ cells enhance oxidative metabolism, raising the ATP/ADP ratio. This metabolic shift closes ATP-sensitive potassium channels (K\textsubscript{ATP}), leading to membrane depolarization~\cite{swisa2015loss}. Depolarization in turn activates voltage-dependent calcium channels (VDCCs), generating a $Ca^{2+}$ influx that triggers insulin granule exocytosis~\cite{swisa2015loss}. A key physiological feature of this process is pulsatile insulin 
secretion, occurring with a period of 4--5 minutes~\cite{shukla2021continuous,laurenti2021measurement,laurenti2020diabetes}. 
Pulsatile secretion is substantially more effective than constant insulin release for maintaining glycemic control, enhancing insulin receptor sensitivity, and preventing desensitization \cite{satin2015pulsatile}.

Importantly, these oscillatory secretions arise from coordinated electrical and metabolic oscillations within the $\beta$-cell network. Disruption of these collective rhythms can lead to insufficient insulin production or impaired insulin responsiveness, contributing to chronic diseases such as type~II diabetes, which is characterized by insulin resistance together with a progressive dysfunction of pancreatic beta cells that persist but exhibit altered functional coordination and secretory activity~\cite{satin2015pulsatile,rutter2015beta}. This underscores the crucial role of intercellular interactions, primarily mediated by gap junctions composed of connexin Cx36 ~\cite{satin2015pulsatile,sherman1991model,ravier2005glucose,loppini2015mathematical}. $\beta$ cells therefore form a dynamic, interconnected network in which coordinated collective behavior is essential for proper islet function~\cite{johnston2016beta}.

Collective dynamics are a hallmark of many biological systems, where coordination among interacting units gives rise to emergent behaviours essential for physiological regulation. In this context, mathematical modelling and numerical simulation provide an essential framework for linking experimental observations to the underlying dynamical mechanisms. Examples include synchronised electrical activity in cardiac tissue~\cite{rohr2004role}, neuronal oscillations that underlie information processing in the brain~\cite{buzsaki2004neuronal,breakspear2017dynamic}, and calcium waves that coordinate intercellular communication in epithelial or glial networks~\cite{dupont2016models}. These systems are often described and analysed using the framework of coupled nonlinear oscillators, which provides insight into how local interactions can generate global temporal patterns such as synchronisation \cite{pikovsky2001universal}, clustering\cite{mcgraw2005clustering}, wave propagation \cite{li2016spiral}, or chimera states\cite{abrams2004chimera}.

Most previous studies have focused on the electrophysiology of $\beta$-cells~\cite{chay1983minimal,bertram2007metabolic,sherman1991model}, successfully describing synchronization via gap junctions and the propagation of electrical waves. However, these models were largely restricted to fast dynamics and did not fully incorporate the slower metabolic processes known to modulate $\beta$-cell excitability. A second generation of studies emphasized that glycolytic and mitochondrial oscillations, together with electrical activity, form an integrated metabolic--electrophysiological system~\cite{bertram2007interaction,merrins2016phase,marinelli2021symbiosis,fridlyand2003modeling}. These frameworks highlighted how the interplay between fast and slow activities regulates the frequency, amplitude, and coordination of islet-wide rhythms. More recently, Sterk \textit{et al.}~\cite{vsterk2023both} proposed an integrated framework explicitly combining fast electrical and slow metabolic oscillators coupled across cells, showing that the interaction of these components shapes the multimodality and functional connectivity of $\beta$-cell networks. Nevertheless, the full range of possible dynamical regimes and their potential relevance for physiological or pathological states remains insufficiently explored.

The present work extends the integrated framework of Sterk~\textit{et al.}~\cite{vsterk2023both} by introducing non-local coupling, inspired by studies such as Shahriari \textit{et al.}~\cite{shahriari2019role}, which showed that non-local interactions can generate rich and spatially structured collective states, including chimera states, in pancreatic $\beta$-cell networks. In their investigations, they considered a non-local network topology based on the Sherman model and mainly investigated the influence of the coupling strength and coupling range on the emergence of stationary chimera patterns. However, within the present investigation, we employ a hybrid Poincaré--Rulkov model that combines slow metabolic oscillations with fast electrical dynamics through a bilayer description of the cellular activity. This framework allows us to investigate the emergence of traveling chimera in pancreatic $\beta$-cell networks. Traveling and non-stationary chimera have previously been reported in generic non-locally coupled networks of excitable systems with electrical and chemical interactions~\cite{mehrabbeik2021synchronization,simo2021traveling} and have been extensively studied as a general phenomenon in nonlinear oscillator networks~\cite{omel2019traveling}. In this paper, we examine these dynamical states in the context of a physiologically motivated pancreatic $\beta$-cell model, thereby extending previous studies toward a more biologically relevant setting. Building upon this perspective, our investigation analyses the spatiotemporal behaviors emerging within a network of coupled $\beta$-cell oscillators integrating both fast electrical activity (modeled by a Rulkov map) and slower metabolic dynamics (represented by a Poincaré oscillator) under a non-local topology. Our goal is to analyze how the coupling structure and the intrinsic properties of individual oscillators shape the spatiotemporal coordination of the network and, consequently, the efficiency of pulsatile insulin secretion.

The remainder of this paper is organized as follows. Sec.~\ref{sec::Mod} introduces the model description. Sec.~\ref{sec::NAD} presents the numerical results and analyzes the collective behaviors exhibited by the network, together with their biological implications. Finally, Sec.~\ref{sec::con} summarizes the main conclusions and discusses the broader significance of our findings.

\section{Mathematical Modeling} \label{sec::Mod}

Let us consider a hybrid multicellular model that combines the fast electrical activity and slow metabolic oscillations of pancreatic $\beta$-cells, following the framework introduced by \v{S}terk \textit{et al.}~\cite{vsterk2023both}. This model should be considered and understood as a  theoretical–phenomenological framework rather than a detailed anatomical reconstruction of a pancreatic islet. 
The theoretical aspect  is related to the fact that it deliberately uses reduced oscillators, and an idealized nonlocal ring topology in order to isolate the dynamical effects of electrical and metabolic coupling. More specifically, the aim of the present model is to clarify how two classes of intercellular interactions — slow metabolic coupling and fast electrical coupling — shape collective $\beta$-cell dynamics. Within this framework, the Poincaré oscillator, which captures the periodic metabolic fluctuations that influence $\beta$-cell excitability is used to  model the slow metabolic dynamics associated with glycolytic and mitochondrial processes that modulate the intracellular ATP/ADP ratio~\cite{bertram2010electrical,marinelli2021symbiosis}. The fast electrical behavior, driven by ion channel dynamics and Ca$^{2+}$ fluxes across the membrane, is represented by the two-dimensional Rulkov map~\cite{rulkov2001regularization}.The phenomenological character of the model lies in the physiological interpretation of its components. Rulkov map is used to represent fast electrical bursting, the Poincaré oscillator accounts for slower metabolic oscillations, and the two coupling pathways mimic, respectively, electrical communication through gap junctions and metabolic exchange between cells.

We consider a network of $N$ cells, where each cell interacts with its neighbors on a nonlocally coupled ring topology through two distinct coupling pathways: an electrical coupling and a metabolic coupling. It is important to recall that the use of a nonlocally coupled ring topology is not intended to reproduce the full three-dimensional architecture of an islet, but rather to provide a minimal and controlled setting in which the interaction range can be systematically varied through the nearest-neighbour parameter. Therefore, this nonlocal framework can be viewed as suitable way to incorporate both short-range gap-junctional interactions and long-range effective communication pathways, such as those mediated by diffusible metabolic or paracrine factors~\cite{benninger2008gap}. The metabolic coupling, characterized by the strength parameter $k_p$, regulates interactions among the slow metabolic oscillations described by the Poincaré oscillator in Cartesian coordinates, as given by Eq.~\ref{eq::PM}.

\begin{widetext}
\begin{equation}
\label{eq::PM}
\left\{ {\begin{array}{*{20}{l}}
{{{\dot x}_j} =  - {y_j}{\omega _j} - \gamma {x_j}(\sqrt {x_j^2 + y_j^2}  - A) + \frac{k_p}{2P}\sum\limits_{i = j - P}^{j + P} {{W_{ji}}} ({x_i} - {x_j})},\\
{{{\dot y}_j} = {x_j}{\omega _j} - \gamma {y_j}(\sqrt {x_j^2 + y_j^2}  - A) + \frac{k_p}{2P}\sum\limits_{i = j - P}^{j + P} {{W_{ji}}} ({y_i} - {y_j})}.
\end{array}} \right.
\end{equation}
\end{widetext} 
%%%
%%%
Here, $\gamma = 1.0$ denotes the relaxation rate and $A = 0.2$ represents the oscillation amplitude. Following Ref.~\cite{vsterk2023both}, the intrinsic frequency $\omega_j$ of each oscillator is randomly drawn from a uniform distribution centered around the average frequency $\omega_{\mathrm{avg}} = 0.006$, with a variability of $\pm 15\%\omega_{avg}$. This heterogeneity reflects differences in the metabolic activity of individual $\beta$-cells, which influence their intrinsic oscillation periods. The term $W_{ji}$ corresponds to the elements of the adjacency matrix, where $W_{ji} = 1$ if node $j$ is connected to node $i$ i.e., when $|i-j|\le P$, and $W_{ji} = 0$ otherwise, thereby defining the structural topology of intercellular coupling within the simulated islet network.

Electrical coupling, governed by the parameter $k_r$ in the two-dimensional Rulkov map (Eq.~\ref{eq::RM}), models the gap-junctional exchange of ions between neighboring $\beta$-cells, where the fast electrical dynamics are described by the variables $u_j$ (membrane potential) and $v_j$ (slow gating variable).
%%%
\begin{widetext}
\begin{equation}
\label{eq::RM}
\left\{ \begin{array}{l}
{u_j}(n + 1) = \frac{{{\alpha _j}(n)}}{{1 + {u_j}{{(n)}^2}}} + {v_j}(n) + \rho{\xi _j}(n) + \frac{k_r}{2P}\sum\limits_{i = j - P}^{j + P} {{W_{ji}}} ({u_i}(n) - {u_j}(n)),\\
{v_j}(n + 1) = {v_j}(n) - \sigma_j {u_j}(n) - \chi_j, 
\end{array} \right.
\end{equation}
\end{widetext}
%%%
{$\xi_j$ is the Gaussian white noise, with zero mean and unit variance, which is included to account} for the stochastic nature of $\beta$-cell activity. Its strength is scaled by the noise amplitude $\rho $. To capture electrophysiological heterogeneity, the parameters $\sigma_j$ and $\chi_j$ are randomly assigned such that each $\sigma_j$ is drawn from the interval $[1,\,1.4]\times10^{-3}$, with $\chi_j = \sigma_j$; see Refs.~\cite{vsterk2023both,stovzer2019heterogeneity} for details.
%%%%
The coupling between the two oscillatory components of each $\beta$-cell, namely the Poincaré oscillator describing the slow metabolic dynamics and the Rulkov map describing the fast electrical activity, is introduced through the excitability parameter $\alpha_j$, defined in Eq.~\ref{eq::alp}. This coupling establishes a unidirectional influence of the fast electrical dynamics on the slow metabolic dynamics. In this formulation, the metabolic variable $x_j$ of the $j$th Poincaré oscillator modulates the excitability of the corresponding Rulkov oscillator, thereby providing a phenomenological link between slow metabolic oscillations and fast membrane-potential bursting.

\begin{equation}
\label{eq::alp}
\alpha_j = \frac{(x_j + A)(\alpha_{\max} - \alpha_{\min})}{2A} + \alpha_{\min},
\end{equation}
where $\alpha_{\min} = 1.95$ and $\alpha_{\max} = 1.995$. As $x_j$ increases, $\alpha_j$ rises accordingly, reflecting enhanced cellular excitability and a higher oscillation frequency.

Although metabolic and electrical dynamics are based on distinct mechanisms, they interact to modulate  intracellular concentrations of calcium and ATP which are the key elements causing insulin secretion. A simplified representation of this interaction highlights their functional interdependence, providing a more coherent description of the overall state of the cells. This interdependence is mathematically modelled by the composite signal introduced in Eq.\ref{eq::com}, which unifies the two dynamics within a coupled framework~\cite{bertram2007metabolic,vsterk2023both}.
\begin{equation}
\label{eq::com}
{c_j} = {x_j} + b{u_j},
\end{equation}
where $b = 0.5$ is used as a weighting coefficient applied to the Rulkov component, and controls the relative contribution of the electrical activity to the composite signal. The proposed model could also be naturally interpreted as a bilayer network in which each pancreatic $\beta$-cell is represented by two coupled dynamical subsystems. The first layer consists of Poincaré oscillators describing the slow metabolic dynamics, whereas the second layer consists of Rulkov maps representing the fast electrical activity. Unless otherwise specified, the parameters of the Poincaré oscillator and Rulkov map are taken from the model proposed by Sterk \textit{et al.}~\cite{vsterk2023both}. For the convenience of the reader, the numerical values of the parameters used in the present study are the same as those reported by Sterk \textit{et al.}~\cite{vsterk2023both} and are listed above. Compared with Ref.~\cite{vsterk2023both}, the main modification concerns the network topology and the definition of the coupling weights $W_{ji}$, which are taken as binary ($W_{ji}\in\{0,1\}$) in the present work rather than distance-dependent.

\section{Numerical analysis and discussion}\label{sec::NAD}

This section is devoted to the numerical analysis of a network of beta cells  described in Sec.~\ref{sec::Mod}, governed by Eq.~\ref{eq::PM} for the slow component and Eq.~\ref{eq::RM} for the fast electrical activity. Since the slow metabolic components are governed by ordinary differential equations, the Poincaré oscillator defined by Eq.~\ref{eq::PM} is numerically integrated using a fourth-order Runge–Kutta algorithm with a time step of $dt=0.05$. By contrast, the fast electrical dynamics are described by the Rulkov map (see Eq.~\ref{eq::RM}) and evolve through discrete-time iterations indexed by $n$. To consistently couple the continuous- and discrete-time components, one iteration of the Rulkov map is taken to correspond to one unit of time and is therefore associated with $1/dt=20$ Runge–Kutta integration steps of the Poincaré oscillator. The simulations are performed over $10^5$ iterations of the Rulkov map, corresponding to $2\times10^6$ integration steps of the Poincaré oscillator. The first $50\%$ of the simulated data are discarded as transient, while the remaining data are retained for the analysis.

\subsection{Phase synchronization in the slow metabolic component}\label{sec::subsec1}

We consider the previously defined network composed of $N = 100$ coupled systems, where each node exhibits both a slow metabolic component and a fast electrical activity. The fast electrical dynamics is explicitly modulated by the underlying metabolic process, reflecting the physiological interplay between cellular metabolism and membrane excitability. A key question here is how the slow metabolic rhythm shapes or influences the collective transitions observed in both the fast electrical and the composite signal of the network. To quantify this collective behavior, and specially the phase synchronization, we employ the order parameter originally introduced by Kuramoto and Battogtokh~\cite{kuramoto2002coexistence} and expressed by Eq.~\ref{eq::kour}:
%%%
%%%
\begin{equation}
   \label{eq::kour}
   R = \left| \frac{1}{N} \sum_{j=1}^{N} e^{i \phi_j} \right|,
\end{equation}
%%%
%%%
where $i^2 = -1$, and $\phi_j$ denotes the phase of the $j^{\mathrm{th}}$ oscillator. The phase can be extracted from the time series of each unit using  the Hilbert transform, or alternatively, computed as $\phi_j = \arctan(y_j/x_j)$ in the case of a two-variable model. 
This order parameter, $R$, quantifies the level of phase coherence among oscillators: $R \approx 1$ indicates  ful phase synchronization, whereas $R \to 0$ corresponds to a fully desynchronized state. Biologically, a high value of $R$ reflects a strong coordination between the metabolic and electrical rhythms across the network, suggesting collective regulation of cellular activity.

\begin{figure}[htp]
        \begin{center}
            \begin{tabular}{c}
            \hspace{-0.5 cm}
                \includegraphics[width=0.55\textwidth]{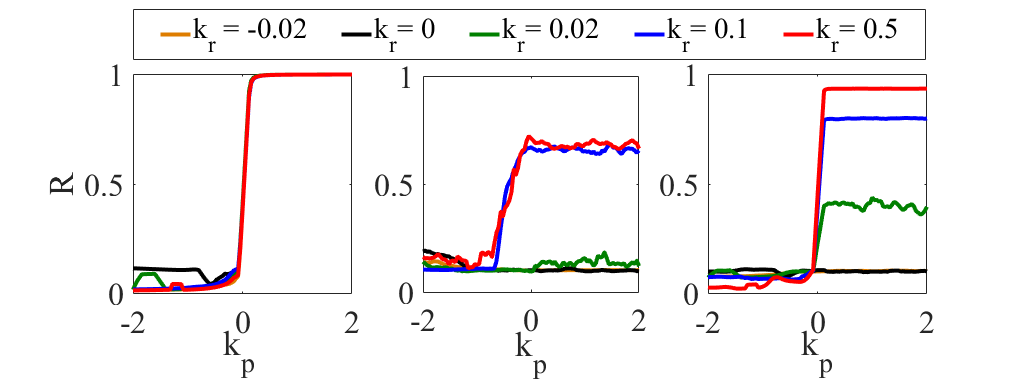}\\
                (a)\hspace{2.3 cm} (b) \hspace{2.3 cm} (c)
            \end{tabular}
            \caption{Phase transition dynamics represented by the order parameter of the (a) metabolic, (b) electrical, and (c) composite signals as a function of the metabolic coupling strength $k_p$, for five values of the electrical coupling strength $k_r = \{-0.02, 0, 0.02, 0.1, 0.5\}$, with a fixed neighborhood size of $P = 20$, and the noise amplitude $\rho=0.005$.}
            \label{fig::Figsync}
        \end{center}
    \end{figure}

Fig.~\ref{fig::Figsync} illustrates the transition to phase synchronization based on the order parameter for the metabolic variables [Fig.~\ref{fig::Figsync}(a)], the electrical activity [Fig.~\ref{fig::Figsync}(b)], and the composite signal [Fig.~\ref{fig::Figsync}(c)] as a function of the metabolic coupling strength $k_p$, for five values of the electrical coupling strength $k_r = \{-0.02, 0, 0.02, 0.1, 0.5\}$ chosen. This result shows that the $\beta$-cell network undergoes a clear transition from incoherent to synchronized dynamics as the metabolic coupling strength $k_p$ increases. In each case, both positive and negative coupling strengths are considered. It should be noted that positive coupling strengths, i.e., $k_p>0$ and $k_r>0$ correspond to attractive interactions that promote synchronization between neighboring cells and therefore represent the physiologically relevant regime. However, negative coupling strengths, i.e., $k_p<0$ and $k_r<0$ correspond to repulsive interactions, that can favor anti-phase or desynchronized collective states and are considered here only as a theoretical extension to explore the full spectrum of collective behaviors supported by the model. Therefore, negative coupling strengths are not intended to represent physiological metabolic or gap-junction interactions.

%%%
%%%
\begin{figure}[htp!]
    \begin{center}
        \begin{tabular}{cc}
            \includegraphics[scale=0.3]{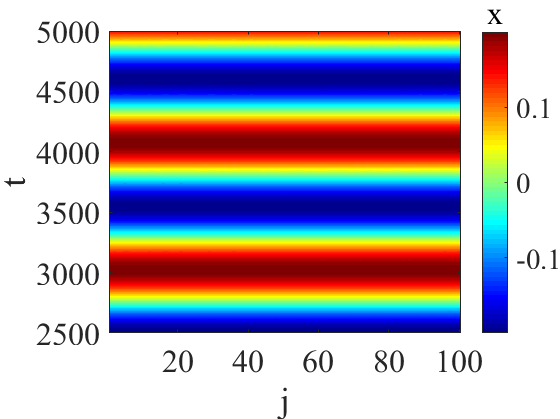} &
            \includegraphics[scale=0.3]{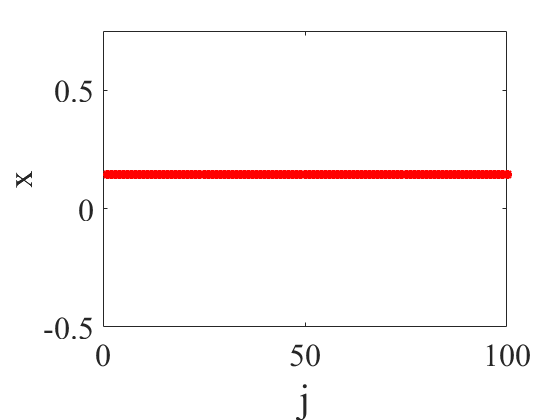}\\
            (a) & (d) \\
            \includegraphics[scale=0.3]{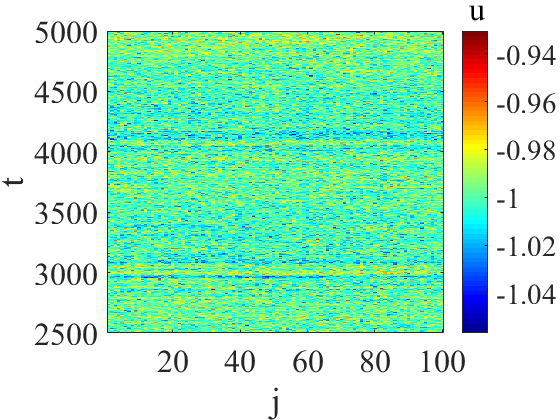} &
            \includegraphics[scale=0.3]{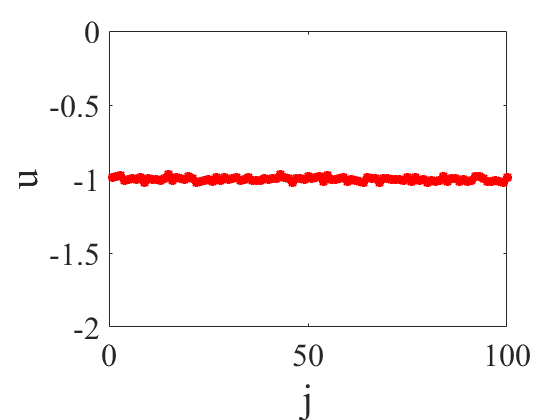}\\
            (b) & (e) \\  
            \includegraphics[scale=0.3]{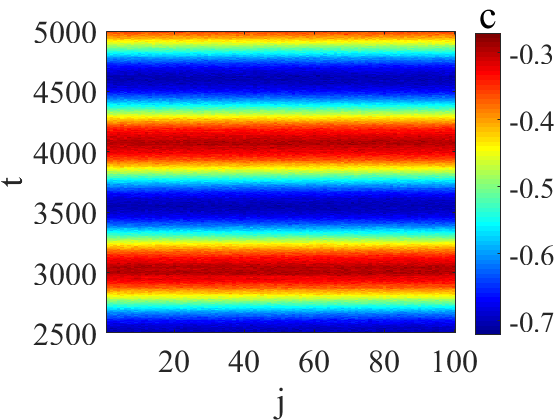} &
            \includegraphics[scale=0.3]{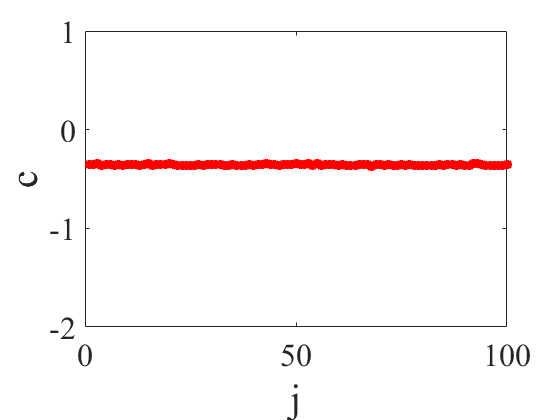} \\
           (c) & (f) \\
        \end{tabular}
        \caption{Spatiotemporal patterns and corresponding snapshots illustrating the synchronization dynamics of (a,d) the metabolic variable $x_i$, (b,e) the membrane potential $u_i$, and (c,f) the composite signal $c_i$. The simulations were performed using $k_p=1$, $k_r=0.1$, $P=20$, and $\rho=0.005$.}
    \label{fig::Figsynctime}
\end{center}
\end{figure}

%%%
%%%
From a biological perspective, the sign of the coupling parameters determines whether the intercellular interactions promote or oppose formation of collective behavior and especially synchronization. As observed in all panels of Fig.~\ref{fig::Figsync}, for negative $k_p$, the order parameter $R$ remains low, indicating desynchronized metabolic and electrical oscillations.  In contrast, when $k_p$ becomes positive, a sharp phase transition occurs, leading to strong collective synchronization of both metabolic and electrical activities, as well as of the composite signal. For illustration, Fig.~\ref{fig::Figsynctime} presents the time series of the metabolic [Fig.~\ref{fig::Figsynctime}(a)], electrical activities [Fig.~\ref{fig::Figsynctime}(b)], and the composite signal [Fig.~\ref{fig::Figsynctime}(c)], together with their corresponding snapshots shown in Fig.~\ref{fig::Figsynctime}(d--f), for the metabolic coupling $k_p = 1$ and electrical coupling $k_r = 0.1$. As predicted by the order parameter in Fig.~\ref{fig::Figsync}, for this set of coupling values the system exhibits fully synchronized behavior in both the slow metabolic and fast electrical variables as well as in the composite signal.

Definitely, positive values of $k_p$ and $k_r$ tend to drive the systems to a scenario that favors the synchronization, and then mimicking the physiological situation in which $\beta$-cells communicate through gap junctions that allow the diffusion of ions (electrical coupling) and small metabolites such as glucose-6-phosphate or ATP/ADP intermediates (metabolic coupling). This positive coupling promotes phase synchronization with coherent bursting in electrical activity and coordinated metabolic oscillations across the islet, thereby supporting pulsatile insulin secretion~\cite{vsterk2023both}. In contrast, negative values of coupling $k_p$ and $k_r$ represent anti-phase or desynchronizing couplings, which leads to the destabilization of collective activity. 
As observed in all panels of Fig.~\ref{fig::Figsync}, except for the metabolic variable shown in Fig.~\ref{fig::Figsync}(a), the degree of electrical coupling $k_r$ strongly modulates the transition to synchronization, with larger positive values ($k_r > 0$) enhancing global synchronization across the network. In contrast, variations in $k_r$ have little influence on the synchronization of the metabolic variable, which is mainly controlled by the metabolic coupling $k_p$ (see Eq.~\ref{eq::PM}). This occurs because electrical coupling through gap junctions primarily coordinates the rapid exchange of ions responsible for fast electrical bursting, whereas the slower metabolic oscillations arise from the diffusive exchange of metabolites that are unaffected by $k_r$.

According to Refs.~\cite{ng2021intercellular,zhang2008cell,vsterk2023both}, experimental  observations indicate that, in pancreatic $\beta$-cell networks, the average number of neighbors of a cell, i.e., the number of connections it maintains through gap junctions, reflects the strength of its coupling with the rest of the network and has a direct biological significance. In our study, this average degree is given by $k = 2P$ due to the symmetry of the ring network. Experimentally, an average of about five to six neighbors per cell, as reported for islets of Langerhans~\cite{zhang2008cell}, corresponds to a normal level of coupling that enables the electrical and metabolic synchronization required for pulsatile insulin secretion. These studies show that a small number of neighbors represents weakly connected cells, which are less synchronized and have only a local influence, whereas a large number of neighbors characterizes highly interconnected cells — so-called “hub cells”— that play a coordinating role in the propagation of electrical and metabolic signals. Therefore, by varying both the communication radius between cells, represented by the neighborhood size $P$, and the level of cellular heterogeneity, the model may help capture part of the functional diversity observed in pancreatic islets. It could also explain how structural connectivity modulates the coherence of the collective activity of $\beta$-cells, particularly how it affects the transition toward synchronization.

\begin{figure*}[htp!]
    \centering
    \resizebox{\textwidth}{!}{%
    \begin{tikzpicture}[
        x=1cm,
        y=1cm,
        axis/.style={
            -{Stealth[length=5pt,width=5pt]},
            line width=1.2pt
        },
        rowlabel/.style={
            font=\small\bfseries,
            anchor=east
        },
        collabel/.style={
            font=\scriptsize,
            anchor=north,
            align=center
        },
        scalelabel/.style={
            font=\scriptsize,
            anchor=west
        }
    ]

        % General dimensions
        \def\colW{8.5}
        \def\boxH{7.25}
        \def\headH{0.85}

        % Enlarged image dimensions
        \def\imgW{7.85}
        \def\imgH{5.45}
        \def\imgX{0.45}
        \def\imgY{1.10}

        % =========================================================
        % Panel A
        % =========================================================
        \begin{scope}

            \draw[
                line width=1.4pt,
                rounded corners=8pt,
                draw=black!70
            ] (0,0) rectangle (\colW,\boxH);

            \node[font=\normalfont\bfseries]
                at ({0.5*\colW},{\boxH-0.5*\headH})
                {A. Intrinsic frequency $\omega$};
            \node[
                anchor=south west,
                inner sep=0pt
            ] (imgA) at (\imgX,\imgY) {%
                \includegraphics[
                    width=\imgW cm,
                    height=\imgH cm,
                    keepaspectratio
                ]{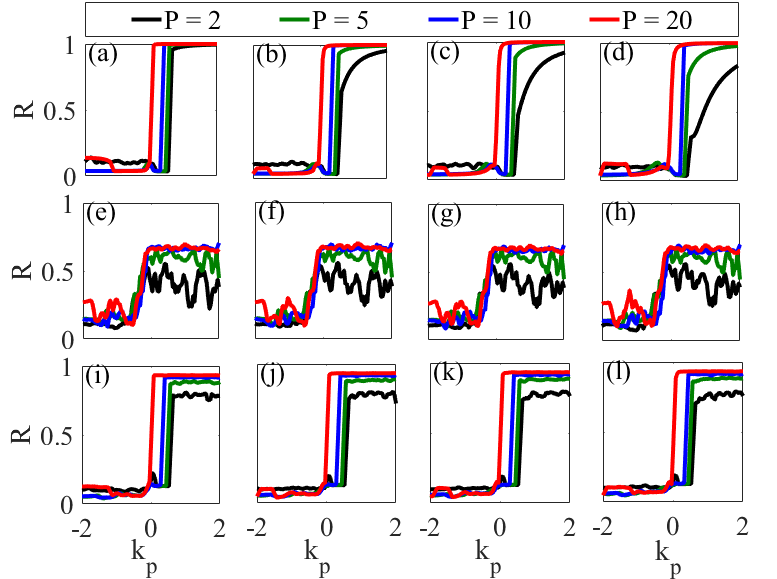}%
            };
            %%%%%%%%
            \begin{scope}[
                shift={(imgA.south west)},
                x={($(imgA.south east)-(imgA.south west)$)},
                y={($(imgA.north west)-(imgA.south west)$)}
            ]
                % Perfectly horizontal and vertical axes
                \draw[axis] (0.00,-0.015) -- (1.0,-0.015);
                \draw[axis] (-0.012,0.04) -- (-0.012,0.965);

                % Row labels
                \node[rowlabel, font=\rowlabelsize]
                at (-0.005,0.845) {MS};
        
                \node[rowlabel, font=\rowlabelsize]
                    at (-0.005,0.565) {ES};
            
                \node[rowlabel, font=\rowlabelsize]
                    at (-0.005,0.285) {CS};

                % Column labels
                \node[collabel] at (0.125,-0.045)
                    {$
                    \pm5\%\,\omega_{\mathrm{avg}}$};

                \node[collabel] at (0.375,-0.045)
                    {$
                    \pm15\%\,\omega_{\mathrm{avg}}$};

                \node[collabel] at (0.625,-0.045)
                    {$
                    \pm20\%\,\omega_{\mathrm{avg}}$};

                \node[collabel] at (0.875,-0.045)
                    {$
                    \pm30\%\,\omega_{\mathrm{avg}}$};

                %\node[scalelabel] at (1.022,-0.015)
                 %   {$+\omega_{\mathrm{avg}}$};
            \end{scope}
        \end{scope}

        % =========================================================
        % Panel B
        % =========================================================
        \begin{scope}[xshift=8.9cm]

            \draw[
                line width=1.4pt,
                rounded corners=8pt,
                draw=black!70
            ] (0,0) rectangle (\colW,\boxH);

            \node[font=\normalfont\bfseries]
                at ({0.5*\colW},{\boxH-0.5*\headH})
                {B. Electrophysiological heterogeneity $\sigma$};

            \node[
                anchor=south west,
                inner sep=0pt
            ] (imgB) at (\imgX,\imgY) {%
                \includegraphics[
                    width=\imgW cm,
                    height=\imgH cm,
                    keepaspectratio
                ]{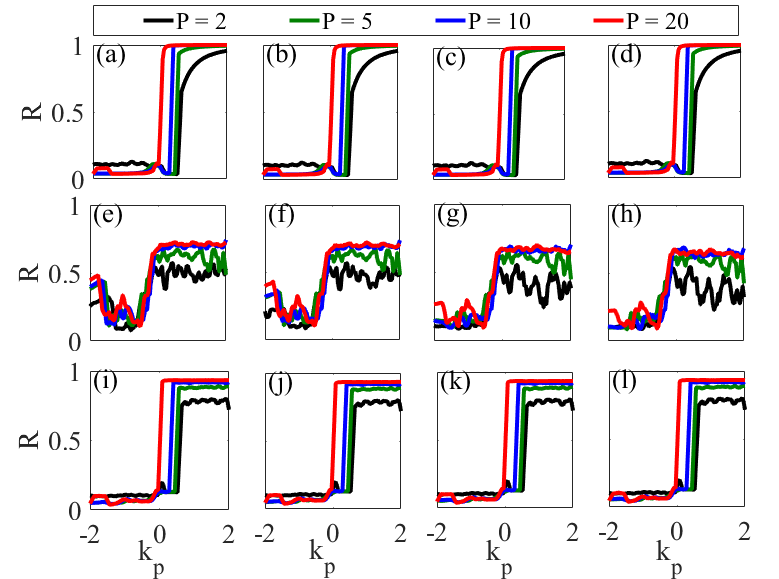}%
            };
            % Normalized coordinate system attached to the image
        
        \begin{scope}[
            shift={(imgB.south west)},
            x={($(imgB.south east)-(imgB.south west)$)},
            y={($(imgB.north west)-(imgB.south west)$)}
        ]
        
            % Perfectly horizontal and vertical axes
            \draw[axis] (0.00,-0.015) -- (1.0,-0.015);
            \draw[axis] (-0.008,0.04) -- (-0.01,0.965);
        
            % Row labels
            \node[rowlabel, font=\rowlabelsize]
                at (-0.005,0.845) {MS};
        
            \node[rowlabel, font=\rowlabelsize]
                at (-0.005,0.565) {ES};
        
            \node[rowlabel, font=\rowlabelsize]
                at (-0.005,0.285) {CS};
        
            % Column labels
            \node[collabel] at (0.15,-0.045)
                {$[1.15,1.25]$};
        
            \node[collabel] at (0.4,-0.045)
                {$[1.10,1.30]$};
        
            \node[collabel] at (0.65,-0.045)
                {$[1.00,1.40]$};
        
            \node[collabel] at (0.9,-0.045)
                {$[0.95,1.45]$};
        
            \node[scalelabel] at (1.01,-0.015)
                {$\times 10^{-3}$};
        \end{scope}
        \end{scope}
    \end{tikzpicture}%
    }
    \caption{Phase-transition dynamics characterized by the order parameters of the metabolic signal (MS), electrical signal (ES), and composite signal (CS) as functions of the metabolic coupling strength $k_p$, for four neighborhood sizes, $P\in\{2,5,10,20\}$ corresponding to the first, second, third, and fourth columns, respectively. The first, second, and third rows correspond to the MS, ES, and CS, respectively. \textbf{(A)} Effect of heterogeneity in the intrinsic metabolic
    frequencies $\omega$, with distribution widths of $\pm5\%$, $\pm15\%$, $\pm20\%$, and $\pm30\%$ around $\omega_{\mathrm{avg}}=0.006$. \textbf{(B)} Effect of electrophysiological heterogeneity through the parameter $\sigma$, whose values are drawn from the intervals $[1.15,1.25]\times10^{-3}$, $[1.10,1.30]\times10^{-3}$, $[1.00,1.40]\times10^{-3}$, and $[0.95,1.45]\times10^{-3}$.  In both panels, moving from left to right corresponds to a progressive increase in the width of the distribution of the considered parameter, namely $\omega$ in panel (A) and $\sigma$ in panel (B), thereby also reflecting an increase in the heterogeneity. Unless otherwise stated, the electrical coupling strength is fixed at $k_r=0.5$, with the noise amplitude $\rho=0.005$.}
    
    \label{fig::figcomp}
\end{figure*}

Fig.~\ref{fig::figcomp} illustrates how local dynamical heterogeneity and structural connectivity influence the transition toward synchronization as the metabolic coupling strength $k_p$ is varied, with the electrical coupling strength fixed at $k_r=0.5$. Local heterogeneity is introduced through the intrinsic metabolic frequency $\omega$ in panel~A and the electrophysiological parameter $\sigma$ in panel~B. The parameter $P$ characterizes the communication radius between $\beta$-cells, i.e., the number of nearest neighboring cells on each side with which a given cell can directly interact. In each panel, i.e., Fig.~\ref{fig::figcomp}A and Fig.~\ref{fig::figcomp}B, the four curves correspond to $P=2$ (black), $P=5$ (green), $P=10$ (blue), and $P=20$ (red), representing progressively larger communication ranges and consequently, increasing local connectivity. From top to bottom, the first, second, and third rows show the synchronization behavior of the metabolic, electrical, and composite signals, respectively.
%%%%%%%%%%
To investigate metabolic heterogeneity in panel~A, the intrinsic frequency $\omega_j$ of each oscillator $j$ is independently drawn from a uniform distribution centered on the mean frequency $\omega_{\mathrm{avg}}=0.006$. Four distribution ranges are considered:
$\omega_{\mathrm{avg}}\pm5\%\omega_{\mathrm{avg}}$ in
Figs.~\ref{fig::figcomp}A(a,e,i), $\omega_{\mathrm{avg}}\pm15\%\omega_{\mathrm{avg}}$ in
Figs.~\ref{fig::figcomp}A(b,f,j),
$\omega_{\mathrm{avg}}\pm20\%\omega_{\mathrm{avg}}$ in
Figs.~\ref{fig::figcomp}A(c,g,k), and
$\omega_{\mathrm{avg}}\pm30\%\omega_{\mathrm{avg}}$ in
Figs.~\ref{fig::figcomp}A(d,h,l). Moving from left to right therefore corresponds to a progressive broadening of the frequency distribution and, consequently, to an increase in metabolic heterogeneity.
%%%%%%%%%%%%%%
A similar procedure is applied to the electrical signal in panel~B. For each oscillator, the parameter $\sigma_j$ is independently sampled from one of the following uniform intervals:
$[1.15,1.25]\times10^{-3}$ in Figs.~\ref{fig::figcomp}B(a,e,i),
$[1.10,1.30]\times10^{-3}$ in Figs.~\ref{fig::figcomp}B(b,f,j),
$[1.00,1.40]\times10^{-3}$ in Figs.~\ref{fig::figcomp}B(c,g,k), and
$[0.95,1.45]\times10^{-3}$ in Figs.~\ref{fig::figcomp}B(d,h,l) with a mean fixed at the center of each interval. Here again, the progressive widening of the sampling interval from left to right reflects an increasing degree of electrophysiological heterogeneity. This investigation shows that increasing the metabolic coupling strength $k_p$ generally promotes the emergence of synchronization in the metabolic, electrical, and composite signals. For the metabolic signal, the order parameter $R$ undergoes a relatively sharp transition from low to high values, indicating a phase transition from a desynchronized to a synchronized state. The critical coupling strength associated with this transition strongly depends on the communication radius $P$. Specifically, larger values of $P$ shift the onset of synchronization toward lower values of $k_p$, whereas smaller communication radii, particularly the case $P=2$, require stronger metabolic coupling and exhibit a more gradual transition. This effect becomes more pronounced as the level of metabolic heterogeneity increases. Interestingly, this analysis suggests that the influence of the electrophysiological parameter $\sigma$ is less pronounced than that of the intrinsic metabolic frequency $\omega$. From this observation, it appears that the synchronization process is more sensitive to heterogeneity in the metabolic dynamics than to variability in the electrophysiological parameter. The electrical signal (see the second rows of Fig.~\ref{fig::figcomp}) differs from the metabolic signal in that its order parameter saturates at an intermediate value rather than approaching full phase synchronization. Consequently, the composite signal closely follows the synchronization transition of the metabolic signal while remaining slightly below unity because of the partial coherence of the electrical dynamics. Globally, these results show that increasing the communication radius, i.e., $P$ enhances the robustness of collective synchronization
against both metabolic and electrophysiological heterogeneity.\\
%%%%%%
Biologically, this means that in pancreatic islets, a higher number of connections between $\beta$-cells through gap junctions facilitates the propagation of both electrical and metabolic signals, thereby improving the spatial and temporal coherence of collective activity and ensuring coordinated pulsatile insulin secretion~\cite{zhang2008cell,vsterk2023both}. In contrast, drastically decreasing the number of intercellular connections, leads to impaired synchronization, as observed under diabetic or stress conditions, and then increased metabolic heterogeneity, and disrupted calcium oscillations and insulin release~\cite{rutter2015beta,gosak2022ca2+}.

\begin{figure*}[htp!]
\centering
\begin{tikzpicture}
\node[anchor=south west, inner sep=0] (tab) at (0,0) {
\begin{tabular}{c}
\hspace{-1.50cm}
\includegraphics[width=1.15\textwidth]{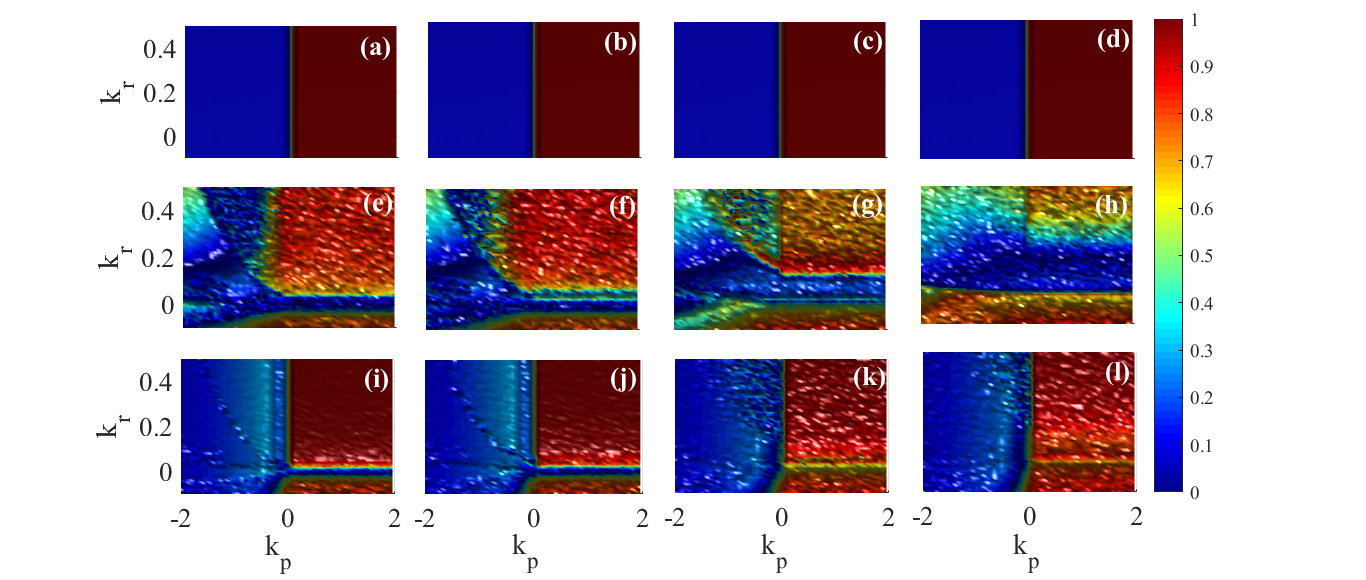}
\end{tabular}
};
\hspace{-.15cm}
\begin{scope}[x={(tab.south east)}, y={(tab.north west)}]
\draw[-{Stealth[length=7pt,width=8pt]}, line width=3pt](0.05,-0.00) -- (0.880,-0.00);
\draw[-{Stealth[length=7pt,width=8pt]}, line width=3pt](-0.0035,0.20) -- (-0.0035,0.950);
\node[anchor=north] at (0.12,0.0)  {$\rho =0.005$};
\node[anchor=north] at (0.34,0.0)  {$\rho =0.008$};
\node[anchor=north] at (0.55,0.0)  {$\rho =0.03$};
\node[anchor=north] at (0.75,0.0)  {$\rho =0.06$};
\node[anchor=east] at (-0.0035,0.850) {MS};
\node[anchor=east] at (-0.0035,0.570) {ES};
\node[anchor=east] at (-0.0035,0.30) {CS};
\end{scope}
\end{tikzpicture}
\caption{
Order parameter in the $(k_p,k_r)$ parameter plane for the metabolic signal (MS), electrical signal (ES), and composite signal (CS), with a fixed neighborhood size $P=20$. 
The first row corresponds to the MS, panels~(a--d); while, the second row to the ES, panels~(e--h); and the third row to the CS, panels~(i--l). 
From left to right, the columns correspond to increasing noise amplitudes, $\rho  =0.005$, $0.008$, $0.03$, and $0.06$, respectively. 
The colorbar indicates the degree of phase synchronization, with low values indicating desynchronized dynamics and values close to one indicating phase synchronization.
}
\label{fig::figkrkp}
\end{figure*}

To illustrate the influence of both metabolic and electrical coupling strengths on the transition to synchronization within this $\beta$-cell network, Fig.~\ref{fig::figkrkp} shows the order parameter in the $(k_p,k_r)$ parameter plane. The three rows correspond respectively to the metabolic signal (MS; panels~(a--d)), the electrical signal (ES; panels~(e--h)), and the composite signal (CS; panels~(i--l)), while the four columns correspond to increasing noise amplitude, $ \rho=0.005$, $0.008$, $0.03$, and $0.06$ from left to right. The neighborhood size is fixed at $P=20$.
Panels~(a--d) show that metabolic synchronization depends almost exclusively on the metabolic coupling strength $k_p$: the order parameter exhibits an abrupt transition from a desynchronized state for $k_p<0$ to a fully synchronized regime for $k_p>0$, without any influence of the electrical coupling $k_r$. Moreover, this behavior remains robust with respect to the noise amplitude over the range of $\rho $ values considered. As mentioned in the comments of Fig.~\ref{fig::Figsync}, biologically, this reflects the essential role of metabolic communication in coordinating $\beta$-cell activity: metabolic coupling regulates the intercellular exchange of metabolites and controls the slow oscillatory processes that underlie insulin secretion. In contrast, the order parameter associated with the electrical variables shown in panels~(e--h) exhibits a much richer patterns. Phase synchronization emerges only when both couplings are sufficiently large, revealing a region of partial synchrony for intermediate values of $k_p$ and $k_r$, surrounded by desynchronized and fully synchronized regimes.However, unlike the metabolic variable, this electrical synchronization remains sensitive to noise, as illustrated in panels~(g,h), where the desynchronized region becomes progressively larger as the noise amplitude increases. This complexity reflects the high sensitivity of electrical activity to the interplay between fast electrical coupling and the slower metabolic feedback, and this behavior arises from the fact that electrical activity in $\beta$-cells depends simultaneously on fast ionic coupling through gap junctions and on the slower metabolic processes that modulate membrane excitability. The dynamics of the composite signal illustrated in panels~(i-l) combine features of both subsystems: its synchronized region closely follows the transition in $k_p$, but is modulated by the electrical coupling, producing an intermediate band of partial coherence similar to that observed in electrical signal for low noise amplitudes. 
Since the composite variable for each node $j$ is defined as $c_j=x_j+b\,u_j$, the difference between two nodes $i$ and $j$ can be written as $c_i-c_j=(x_i-x_j)+b(u_i-u_j)$.
When the metabolic subsystem is synchronized as in the red domain of the first row, $x_i-x_j \approx 0$ for all pairs of cells. Consequently, the contribution of the electrical variable to the spatial dispersion of $c_j$ is effectively weighted by the factor $b<1$. As a result, the coherent or synchronized metabolic component acts as an organizing background that can preserve phase coherence in the composite signal, even in parameter regions where the electrical signal alone does not exhibit global phase synchronization. From a biological viewpoint, this result suggests that coordinated metabolic activity may contribute to preserving islet-wide functional coherence even when electrical communication is impaired, thereby supporting robust pulsatile insulin secretion under heterogeneous or perturbed conditions.

 \subsection{Traveling chimera and traveling wave}\label{sec::subsec2}

Beyond phase synchronization, coupled $\beta$-cell networks are known to display a very rich spectrum of spatiotemporal collective behaviors driven by the interplay between electrical and metabolic interactions. From a biological perspective, pancreatic islets depend on fast electrical coupling through connexin-36 gap junctions to coordinate calcium oscillations~\cite{zhang2008cell}, while metabolic activity propagates more slowly via diffusive exchange of metabolites and feedback through oxidative pathways~\cite{vsterk2023both}. In this context, two dynamical regimes are particularly relevant: traveling chimera states~\cite{chow1998traveling} and traveling metabolic waves~\cite{gosak2022ca2+}. 

\begin{figure}[htp!]
     \begin{center}
        \begin{tabular}{cc}
     \includegraphics[scale=0.285]{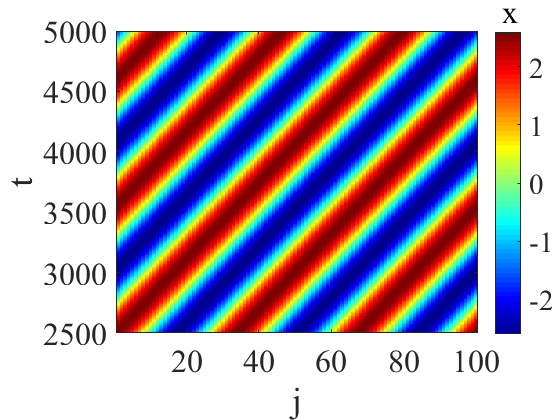} &
    \includegraphics[scale=0.285]{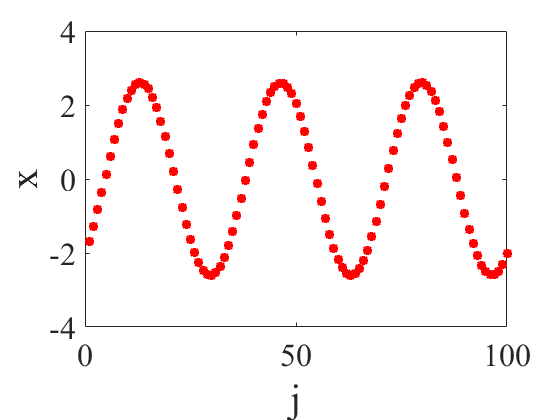}\\
             (a) & (d) \\
    \includegraphics[scale=0.285]{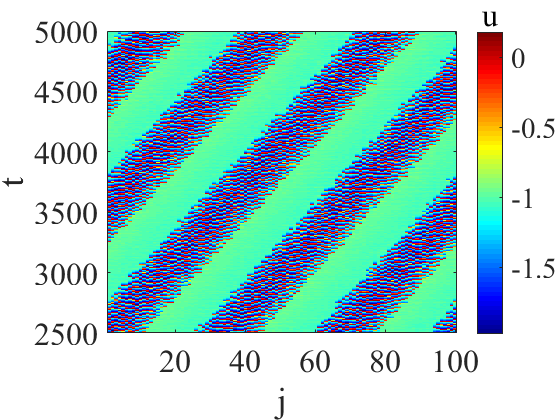}&
     \includegraphics[scale=0.285]{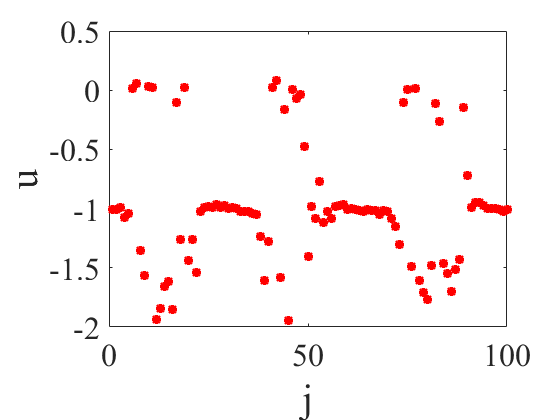}\\
              (b) & (e) \\
    \includegraphics[scale=0.285]{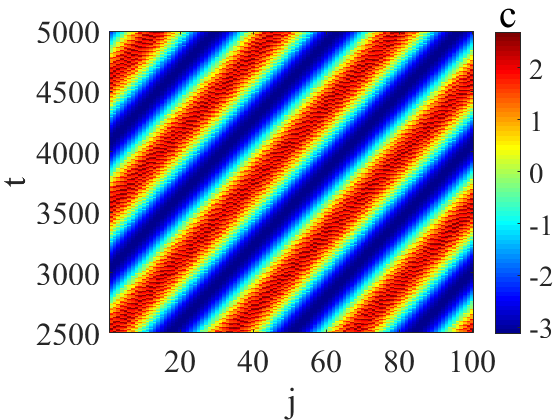}&
    \includegraphics[scale=0.285]{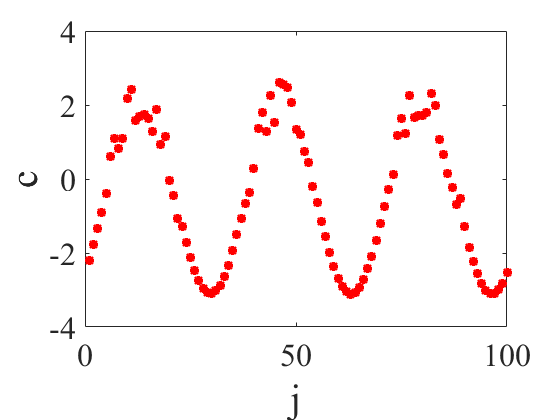} \\
               (c) & (f) 
        \end{tabular}
        \caption{Spatio-temporal and snapshot diagrams illustrating: (a,d) the metabolic variable $x_i$, exhibiting a traveling wave; (b,e) the membrane potential $u_i$, displaying a traveling chimera state; and (c,f) the composite signal $c_i$, showing a weak traveling chimera very close to traveling wave pattern. The parameters used are $k_p = -2$, $k_r = -0.02$, $P = 20$, and $\rho=0.005$}.
    \label{fig::figtctw}
    \end{center}
\end{figure}

Fig.~\ref{fig::figtctw} illustrates the coexistence of distinct spatiotemporal behaviors in the electrical and metabolic variables under the same parameter regime, $k_p = -2$, $k_r = -0.02$, a neighborhood size of $P = 20$ nodes, and a noise amplitude $\rho=0.005$. The temporal evolution of the metabolic variable plotted in Fig.~\ref{fig::figtctw}(a) with a snapshot in Fig.~\ref{fig::figtctw}(b) is organized into a coherent traveling wave, which represents coherent slow fronts that propagate across the tissue, consistent with experimentally reported metabolic and NAD(P)H waves in intact islets~\cite{slaby2009oscillatory,gosak2022ca2+}. Whereas the electrical activity represented in Fig.~\ref{fig::figtctw}(b,e) exhibits a traveling chimera state characterized by the coexistence of coherent and incoherent regions drifting across the ring, reflecting complex partial synchrony in the fast electrical subsystem. The composite signal in Fig.~\ref{fig::figtctw}(c,f) reflects a combination of these two behaviors and reveals a weak propagating chimera very close to a traveling wave pattern shaped by the interaction between slow metabolic coupling and fast electrical variability. Together, these panels highlight the ability of the coupled $\beta$-cell network to sustain qualitatively different dynamical regimes simultaneously, driven by the interplay between electrical and metabolic processes.

\begin{figure}[htp!]
  \begin{center}
   \begin{tabular}{c}
   \hspace{-1 cm}
    \includegraphics[scale=0.38]{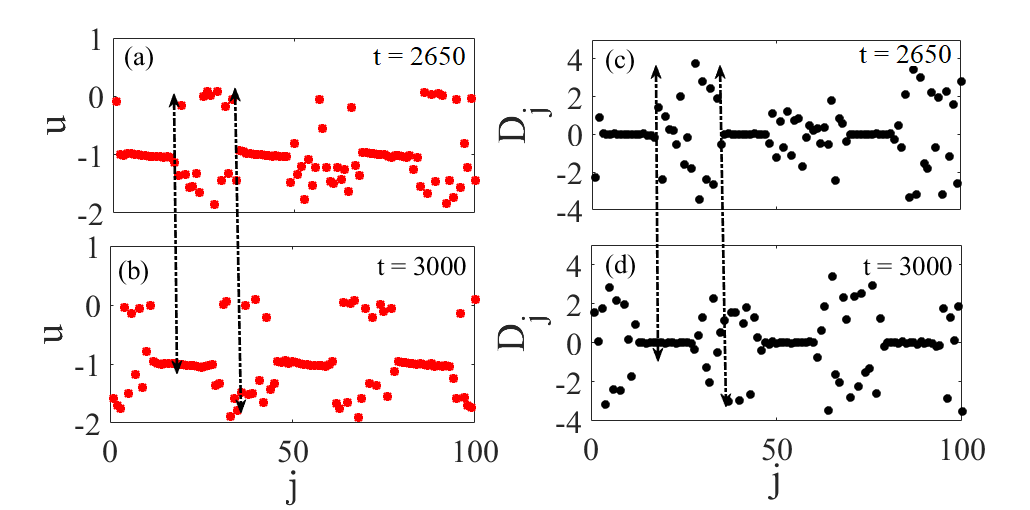} 
   \end{tabular}
   \caption{Snapshots of the electrical variable $u$ ($a,b$) and the corresponding local curvature $D_j$ ($c,d$) at $t=2\,650$ and $t=3\,000$ for $k_r=-0.02$, $k_p=-2$, and $\rho=0.005$. Coherent domains form smooth, continuous segments, whereas incoherent regions show irregular fluctuations. The local curvature sharpens these transitions, with large $D_j$ indicating phase or amplitude disruptions.}
   \label{fig::figlc}
  \end{center}
 \end{figure}

To corroborate the existence of a traveling chimera in the electrical systems $u$, we employ the local curvature measure $D_j(t)$ defined in Eq.~\ref{eq::eqd}, which allows us to distinguish coherent and incoherent regions by identifying points of strong local desynchronization~\cite{mao2025traveling,kemeth2016classification}:
\begin{equation}
    D_j(t) = u_{j+1}(t) - 2u_j(t) + u_{j-1}(t).
    \label{eq::eqd}
\end{equation}
    To demonstrate the non-stationary nature of the chimera, Fig.~\ref{fig::figlc} displays snapshots of the fast electrical variable $u_j$ at two time instants, $t=2\,650$ and $t=3\,000$ (first column), together with the corresponding local curvature profile $D_j(t)$ (second column). At both time instants, the electrical variable exhibits the coexistence of coherent domains with nearly identical phases and incoherent regions displaying irregular fluctuations. However, a clear spatial shift of these domains is observed between the two snapshots, confirming that the boundaries separating coherent and incoherent groups drift over time a defining signature of a traveling chimera~\cite{simo2021traveling}.

Let us introduce the local order parameter~\cite{simo2021traveling} to quantify the degree of synchronization in the neighborhood of each oscillator. Unlike the global order parameter $R$, the local order parameter $L_j$ allows us to track the formation, stability, and motion of coherent clusters (where $L_j \approx 1$) and incoherent domains (where $L_j \approx 0$) along the ring. It is defined in Eq.~\ref{eq::eqlop}.
\begin{equation}
   \label{eq::eqlop}  
  L_j(t) = \left|\frac{1}{2P}\sum_{|j - k| \le P} e^{i\phi_k(t)} \right|, \quad j = 1, 2, \ldots, N,
\end{equation} 
where $i^2=-1$, $P$ denotes the number of neighbors to the left and right of element $j$, and $\phi_k(t)$ is the instantaneous phase of oscillator $k$.\\

\begin{figure*}[htp!]
   \begin{center}
    \begin{tabular}{ccc}
     \includegraphics[scale=0.35]{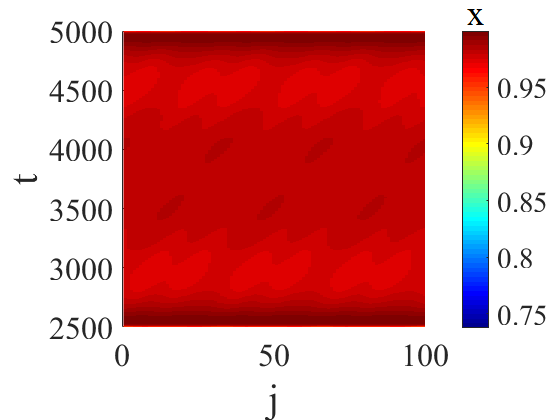}&
     \includegraphics[scale=0.35]{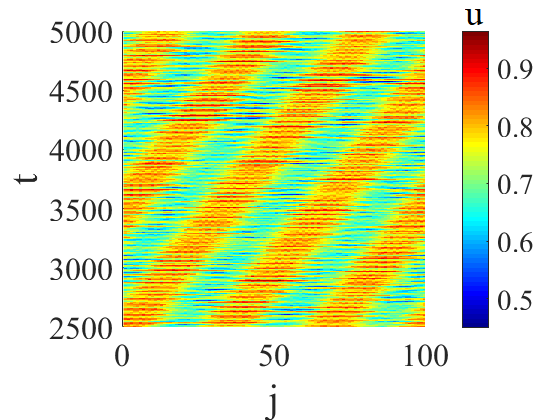}&
     \includegraphics[scale=0.35]{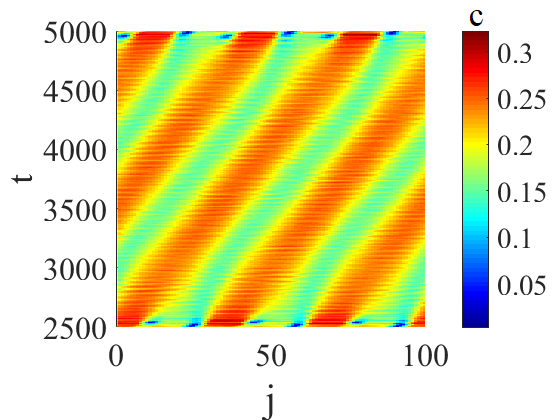}\\
     
          (a) &  (b) &  (c)  
    \end{tabular}
    \caption{Spatio-temporal evolution of the local order parameter associated with the metabolic variable $x$ (a) and the membrane potential $u$ (b), for the same parameters as in Fig.~\ref{fig::figtctw}. The amplitude of $L_j$ is indicated by the color scale.}
    \label{fig::figlop}
   \end{center}
\end{figure*}  

Fig.~\ref{fig::figlop} presents the spatio-temporal evolution of the local order parameter $L_j$ for the metabolic variable $x$ (a), the membrane potential $u$ (b), and the composite signal (c), computed for the same parameters as in Fig.~\ref{fig::figtctw}. Panel~(a) shows that the metabolic system remains almost uniformly synchronized, with $L_j \approx 1$ across space and time. This is in agreement with the fact that the metabolic coupling is strong enough to enforce global coherence. In contrast, the electrical variable displayed in panel~(b) exhibit a clear traveling chimera pattern, where coherent and incoherent regions coexist and drift along the ring, resulting in a characteristic diagonal structure in the $j$–$t$ plane. This non-stationary partial synchrony is further reflected in the composite signal in panel~(c), where the propagation of partially coherent domains is still visible but with reduced amplitude, indicating that the composite dynamics inherit features from both the slow synchronized metabolic system and the fast heterogeneous electrical one.

In the present context, traveling chimera represents a biologically relevant intermediate regime between full synchronization and complete desynchronization. Their coexistence of coherent and incoherent behaviors in the  electrical activity allows the islet to maintain coordinated insulin secretion while preserving functional heterogeneity and adaptability. This spatiotemporal organization may enhance robustness to cellular variability and prevent pathological lock-in to uniform oscillatory modes. From a physiological standpoint, the presence of traveling chimera patterns may therefore support flexible and efficient pulsatile insulin release, whereas their disruption---as observed in diabetic conditions---could lead to impaired signal propagation, loss of oscillatory structure, and ultimately dysfunctional hormone secretion.

To characterize the motion of the incoherent domain in a traveling chimera state, we introduce a velocity measure $v$. This quantity quantifies the drift of the coherent--incoherent boundary along the ring and provides a clear criterion to distinguish stationary or coherent behavior from the traveling chimera. To compute the velocity, we make use of the local curvature defined in Eq.~\ref{eq::eqd}, which allows us to identify the coherent regions characterized by the condition $|D_i| < \delta = 0.01\,D_m$, where $D_m = \max_i |D_i|$~\cite{mao2025traveling}. Thus, the fraction of coherent regions can be expressed as in Eq.~\ref{eq::eqg}.
%%%
\begin{equation}
    g_0(t) = \int_{0}^{\delta} g\!\left(t, |D_j|\right)\, d|D_j|,
    \label{eq::eqg}
\end{equation}
%%%
where $g(t,|D_j|)$ denotes the probability density of the local curvature. Therefore, $g_0=1$ corresponds to a fully coherent state, $g_0\approx0$ to a fully incoherent/desynchronized state, and $0<g_0<1$ characterizes chimera states.\\
%%%%%%
To determine whether the chimera travels, we compare two snapshots separated by a short time interval $\Delta t=t_2-t_1$. Then, the propagation velocity could be estimated from the empirical relation $v=g/\Delta t$, where $g$ denotes the displacement of the coherent domain during the interval $\Delta t$. More specifically, let us denote by $g_1$ and $g_2$ the displacements of the leading and trailing boundaries of the coherent domain, respectively, between the time interval $t_1$ and $t_2$. The corresponding instantaneous velocities are defined as Eq.~\ref{eq::eqdi}

%%%
\begin{equation}
\begin{aligned}
   v_1 &=  \lim_{T \to \infty} \frac{1}{T} \int_{t}^{t+T} v_1(t)\, dt, 
            \quad \text{and} \\
   v_2 &= \lim_{T \to \infty} \frac{1}{T} \int_{t}^{t+T} v_2(t)\, dt,
\end{aligned}
\label{eq::eqdi}
\end{equation}
%%%
The long-time average propagation velocity of the chimera is then computed by Eq.~\ref{eq::eqvm}
\begin{equation}
    v = \frac{v_1 + v_2}{2}.
    \label{eq::eqvm}
\end{equation}
A positive value ($v > 0$) characterizes a traveling chimera, whereas $v \to 0$ may indicate a fully incoherent dynamics, a  stationary chimera, or a fully coherent state. Throughout this investigation, the propagation velocity is expressed in cells per unit integration time of the discrete Rulkov map. Since one iteration of the Rulkov map corresponds to $1/dt=20$ fourth-order Runge--Kutta integration steps of the Poincaré oscillator ($dt=0.05$). Consequently, all reported propagation velocities are computed using a unique and consistent time scale.

\begin{figure*}[htp!]
\centering
\begin{tikzpicture}
\node[anchor=south west, inner sep=0] (tab) at (0,0){
\begin{tabular}{ccc}
\hspace{-1.50cm}
\includegraphics[scale=0.32]{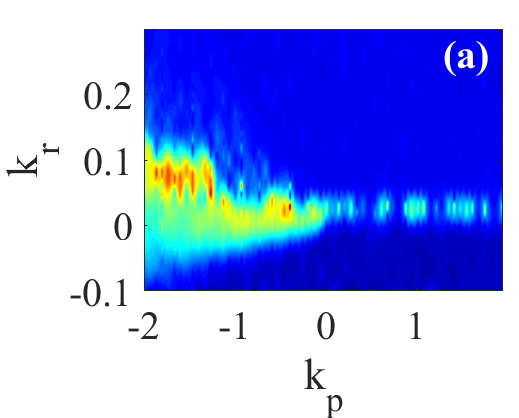}
\includegraphics[scale=0.32]{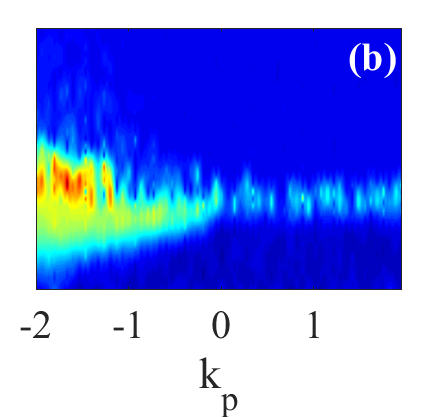}&
\includegraphics[scale=0.32]{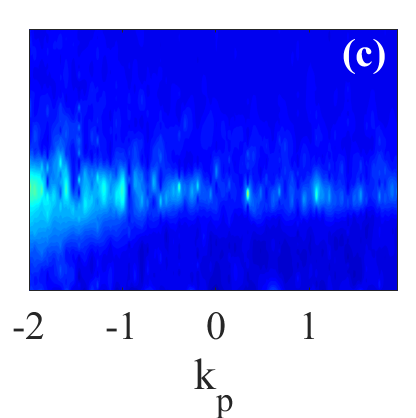}&
%\hspace{.04cm}
\includegraphics[scale=0.32]{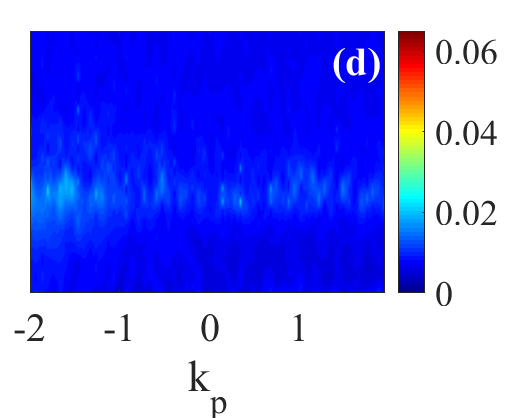}
\end{tabular}
};
\hspace{-.60cm}
\begin{scope}[x={(tab.south east)}, y={(tab.north west)}]
\draw[-{Stealth[length=7pt,width=8pt]}, line width=3pt](0.042,0.0) -- (0.990,0.0);
\node[anchor=north] at (0.12,0.0)  {$\rho  = 0.005$};
\node[anchor=north] at (0.38,0.0)  {$\rho  = 0.008$};
\node[anchor=north] at (0.65,0.0)  {$\rho = 0.03$};
\node[anchor=north] at (0.9,0.0)  {$\rho  = 0.06$};
\end{scope}
\end{tikzpicture}
\caption{Two-dimensional maps illustrating the combined effects of the coupling parameters $k_p$ and $k_r$ on the traveling chimera velocity $v$ for $P=20$, for different values of the noise intensity $\rho$: (a) $\rho=0.005$, (b) $\rho=0.008$, (c) $\rho=0.03$, and (d) $\rho=0.06$. The colorbar indicates the magnitude of the velocity $v$. Nonzero values of $v$ identify parameter regions where traveling chimera states may occur, whereas $v \to 0$ corresponds to coherent or stationary regimes.}
\label{fig::figNoise}
\end{figure*}

Fig.~\ref{fig::figNoise} shows how the traveling chimera velocity $v$ depends on the metabolic and electrical coupling strengths, $k_p$ and $k_r$, and assesses the robustness of traveling chimera propagation against stochastic perturbations for four increasing values of the noise intensity $\rho = 0.005, 0.008, 0.03$, and $0.06$).
The velocity remains close to zero for most of the parameter space (see blue domain), indicating stationary chimera states or fully synchronized dynamics. However, a distinct region of nonzero velocity emerges for weak to moderately negative metabolic coupling ($k_p \lesssim 0$) combined with small positive electrical coupling ($k_r \gtrsim 0$). For low noise amplitude, $\rho = 0.005$ (see panel~(a)) and $\rho = 0.008$ (see panel~(b)), nonzero values of the velocity are observed over a relatively wide region of the $(k_p,k_r)$ parameter plane, suggesting that traveling chimera states can persist under weak to moderate stochastic perturbations (see panel~(c)) within the parameter range considered here. When the noise amplitude is increased, the region supporting nonzero propagation velocity becomes progressively smaller. This reduction is particularly visible for $\rho = 0.06$ (see panel~(d)), where the magnitude of the velocity of the traveling chimera becomes very small, approaching zero. In conclusion, traveling chimera states become progressively weaker as the noise amplitude increases, and sufficiently strong noise may suppress the coherent--incoherent organization required for their emergence.

\begin{figure}[htp!]
   \begin{center}
    \begin{tabular}{c}
     \includegraphics[scale=0.5]{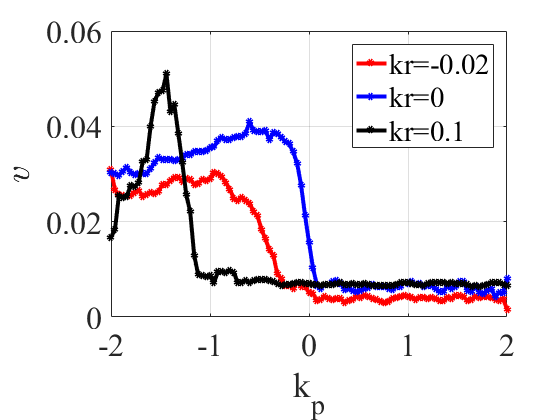}
    \end{tabular}
    \caption{Influence of coupling parameters on chimera motion. The velocity $v$ is shown as a function of the metabolic coupling $k_p$ for three fixed electrical coupling strengths, $k_r = \{-0.02,\,0,\,0.1\}$, with $P = 20$, and $\rho=0.005$}.
    \label{fig::figvkp}
   \end{center}
\end{figure} 

For relatively low or moderate noise amplitude, the transition toward synchronization or coherence emerges as the coupling strengths increase, progressively reducing the mobility of the incoherent domain. In this regime, the incoherent domain drifts along the ring, giving rise to a traveling chimera. For illustration of this transition, Fig.~\ref{fig::figvkp} shows the evolution of the velocity $v$ as a function of the metabolic coupling strength $k_p$ for three values of the electrical coupling strength, $k_r = \{-0.02,\,0,\,0.1\}$ and the noise amplitude $\rho=0.005$. The velocity decreases as either coupling becomes too strong, reflecting how excessive metabolic or electrical synchronization suppresses spatial asymmetry and ultimately inhibits the propagation of the chimera. This behavior is observed for low noise amplitude, $\rho=0.005$ (see Figs.~\ref{fig::figvkp} and~\ref{fig::figNoise}(a)), moderate noise amplitude, $\rho=0.008$ (see Fig.~\ref{fig::figNoise}(b)), and $\rho=0.03$ (see Fig.~\ref{fig::figNoise}(c)). Biologically, this behavior implies that traveling desynchronization waves are more likely to occur when electrical coupling is present but not dominant, and when metabolic interactions do not fully enforce global coherence—conditions that may correspond to heterogeneous or partially impaired $\beta$-cell networks.
This interpretation on traveling chimera is consistent with observation of heterogeneous $\beta$-cell activity, showing that not all cells are active simultaneously, with subsets of cells alternating between active and resting phases~\cite{johnston2016beta}. It also aligns with biophysical studies demonstrating that partially synchronous activity patterns can support pulsatile insulin secretion \cite{benninger2018new}.

\subsection{Emergent Dynamics and Islet Calcium Activity}\label{sec::subsec3}

The collective patterns emerging from our simulations—such as phase synchronization, metabolic waves, and traveling chimera states provide a qualitative framework for interpreting calcium imaging data in pancreatic islets. In particular, the global synchronization of metabolic and electrical oscillations observed in the model reflects the coordinated Ca$^{2+}$ activity experimentally reported between neighboring $\beta$-cells, emphasizing the crucial role of network connectivity and gap junction-mediated coupling in regulating pulsatile insulin secretion \cite{benninger2008gap}.
Similarly, traveling chimera states reproduce qualitative patterns in which some cells remain synchronized while others, phase-shifted, propagate across the islet. Such patterns could correspond to local heterogeneity of Ca$^{2+}$ activity reported in intact islets \cite{gosak2022ca2+}, suggesting that even a simplified model can capture essential spatio-temporal dynamics observed in vivo.\\
These analogies are further  supported by both experimental and theoretical studies. Gap junction-mediated electrical coupling synchronizes Ca$^{2+}$ oscillations, and disruption of coupling decreases the amplitude and propagation speed of calcium waves 
\cite{benninger2008gap, ravier2005loss}. Moreover, dual oscillator models combining slow metabolic and fast calcium/electrical components reproduce key features of $\beta$-cell activity, such as coordinated oscillations and wave propagation across the islet \cite{bertram2007metabolic}. Finally, recent high-resolution calcium imaging studies reveal wave-like propagation and functional sub-networks in human and rodent islets, consistent with heterogeneous activity patterns analogous to the simulated chimera states \cite{gosak2022ca2+}.
Overall, these observations suggest that our simplified phenomenological model captures core mechanisms of collective $\beta$-cell dynamics and provides a mechanistic interpretation of spatio-temporal patterns observed in calcium imaging, even if quantitative agreement with specific experimental data is not intended.

\section{Conclusion}\label{sec::con}

This work proposed and analyzed the collective dynamics of a pancreatic $\beta$-cell network by coupling Rulkov oscillators for fast electrical activity  with Poincaré oscillators describing slow metabolic processes, and examined their behavior under a nonlocal coupling topology. Such a topology is biologically meaningful, as $\beta$-cells within an islet do not interact only with their direct neighbors through gap junctions, but also effectively over longer distances via diffusible metabolic factors.  This hybrid formalism model provides a more physiologically grounded description of $\beta$-cell behavior, surpassing traditional approaches that treat intrinsic dynamics and coupling mechanisms independently. The goal of this work was to elucidate how the interplay between electrical and metabolic coupling shapes the emergent spatiotemporal activity patterns within the islet.

The analysis revealed a rich repertoire of dynamical regimes, including phase synchronization, traveling waves, and traveling chimera states, which are hybrid patterns in which coherent and incoherent domains coexist and propagate through the network. Such diversity reflects the competing influences of electrical and metabolic coupling, which jointly shape the balance between local coordination and global heterogeneity. Quantitative diagnostics such as the (global) order parameter , the local order parameter, spatiotemporal diagrams, and velocity measurements shows that modulation of the coupling strengths induces sharp transitions between these regimes. These results highlight the sensitivity of $\beta$-cell populations to intercellular coupling and illustrate how small variations in communication pathways can reorganize collective behavior, potentially impacting the efficiency and/or robustness of insulin secretion.

Traveling chimera states, in particular, provide a powerful conceptual framework for interpreting the spatiotemporal organization of islet activity. The coexistence of coherent and incoherent domains in this traveling chimera, combined with their ability to propagate across the network, offers a natural mechanism for distributing metabolic information while avoiding the limitations of global synchrony. Such regimes may help explain how pancreatic islets maintain robust pulsatile insulin secretion while retaining sufficient flexibility to respond to fluctuating glucose levels. More broadly, traveling chimeras illustrate how partial coordination—rather than complete synchrony—can enhance the efficiency 
and adaptability of collective behavior in biological systems.

The findings of this investigation are broadly consistent with experimental observations and provide a phenomenological perspective on collective dynamics in $\beta$-cell networks. While the model is not intended as a detailed biophysical description, it highlights the potential importance of intercellular interactions in shaping coordinated activity and pulsatile insulin secretion. More generally, the results suggest possible mechanisms through which alterations in collective dynamics may contribute to impaired glucose regulation under pathological conditions.

\section*{Acknowlegments}
C.S. warmly thanks Aimé C. MOMO for the fruitful discussions. T.N. acknowledges support from the ``Reconstruction, Resilience and Recovery of Socio-Economic Networks'' RECON-NET - EP\_FAIR\_005 - PE0000013 ``FAIR'' -
PNRR M4C2 Investment 1.3, financed by the European Union
– NextGenerationEU.

\bibliography{references}
\end{document}